\documentclass[aps,prb,twocolumn,superscriptaddress,showpacs,floatfix,longbibliography,10pt]{revtex4-2}
\usepackage[utf8]{inputenc}
\usepackage{geometry}
\usepackage{amsmath,amsthm,amsfonts,amssymb,amscd}
\usepackage{multirow,booktabs}
\usepackage[table]{xcolor}
\usepackage{fullpage}
\usepackage{lastpage}
\usepackage{enumitem}
\usepackage{fancyhdr}
\usepackage{mathrsfs}
\usepackage{physics}
\usepackage{wrapfig}
\usepackage{bbm}
\usepackage{braket}
\usepackage[retainorgcmds]{IEEEtrantools}

\usepackage{hyperref}
\hypersetup{
  colorlinks   = true,    % Colours links instead of ugly boxes
  linkcolor    = blue,    % Colour of internal links
  citecolor    = blue     % Colour of citations
}
\usepackage{url}            % simple URL typesetting
\usepackage{empheq}
\usepackage{framed}
\usepackage{xcolor}
\usepackage{comment}
\usepackage{stmaryrd}

\def\hpsi{\hat{\psi}}

\def\d{{\rm d}}

\def\yt{{\tilde y_{\tau,k} }}
\def\yo{{\tilde y_{0,k} }}

\def\yi{{\tilde y_{i,k} }}
\def\yj{{\tilde y_{j,k} }}
\def\yn{{\tilde y_{n,k} }}

\begin{document}

\title{Functional renormalization group study of a dissipative Bose--Hubbard model}

\author{Oscar Bouverot-Dupuis}
\thanks{All authors contributed equally to this work}
\affiliation{Universit\'{e} Paris Saclay, CNRS, LPTMS, 91405, Orsay, France}
\affiliation{IPhT, CNRS, CEA, Universit\'{e} Paris Saclay, 91191 Gif-sur-Yvette, France}

\author{Vincent Grison}
\thanks{All authors contributed equally to this work}
\affiliation{Sorbonne Universit\'{e}, CNRS, Laboratoire de Physique Th\'{e}orique de la Matière Condens\'{e}e, LPTMC, F-75005 Paris, France}

\author{Nicolas Paris}
\thanks{All authors contributed equally to this work}
\affiliation{Sorbonne Universit\'{e}, CNRS, Laboratoire de Physique Th\'{e}orique de la Matière Condens\'{e}e, LPTMC, F-75005 Paris, France}
\affiliation{Universit\'{e} Paris Cit\'{e}, CNRS, Laboratoire Mat\'{e}riaux et Ph\'{e}nomènes Quantiques, 75013 Paris, France}

\begin{abstract}
We investigate the phase diagram of a one-dimensional dissipative Bose–Hubbard model using the nonperturbative functional renormalization group (FRG). Each lattice site is coupled to an independent bath, generating long-range temporal interactions that encode non-Markovian dissipation. For a broad class of bath spectra—ohmic, sub-ohmic, and super-ohmic—we identify two competing low-energy regimes: a Luttinger-liquid line of fixed points and a dissipative fixed point characterized by finite compressibility, vanishing superfluid stiffness, and universal scaling exponents, separated by a Berezinskii–Kosterlitz–Thouless transition. The FRG framework is essential here, as it provides access to the complete renormalization group flow and all fixed points from a single microscopic action, beyond the reach of perturbative or variational methods. This work establishes a unified and systematically improvable framework for describing dissipative quantum phases in one dimension.
\end{abstract}

%\date{\today}

\maketitle

\tableofcontents

\section{Introduction}
Dissipative effects in quantum systems capture the influence of an external environment, or bath, on a system of interest. Such a distinction between system and bath arises naturally in Bose–Fermi mixtures, where the fermionic component typically modifies the properties of the bosonic subsystem~\cite{Gunter_2006,best2009}. This separation becomes all the more pronounced in mixed-dimensional setups, where species-specific optical lattices confine the two components to different dimensionalities~\cite{LeBlanc_2007,Lamporesi_2010}. Similar physics can be realized in solid-state platforms such as arrays of shunted Josephson junctions coupled to transmission lines acting as dissipative environments~\cite{chakravarty1988}.

From a theoretical perspective, quantum dissipation can be modeled by coupling a system to a macroscopic number of environmental degrees of freedom \cite{Caldeira_1983,Leggett_1987,Weiss_2012}. By considering a simple bath made of harmonic oscillators, one can exactly integrate it out and produce an effective retarded, i.e. long-range in time, system-system interaction. This framework accounts for non-Markovian dissipative effects which cannot be studied with standard techniques such as the Lindblad master equation. Initially developed for single-degree-of-freedom systems \cite{Schmid_1983,Fisher_1985}, this approach has since been extended to study various system–bath couplings in many-body systems \cite{Castro_Neto_1997,Artemenko_2007,Weber_2022,Min2024}. Note that for one-dimensional systems, dissipative effects are akin to long-range interactions in space \cite{feng2023,Orignac_2025} after an exchange of space and time.

Focusing on one-dimensional quantum systems with a local bath coupled to each site, dissipation can arise in two distinct forms: coupling either to the phase of the field operator or to the density fluctuations. In the former case, dissipation can enhance superfluidity, leading to true long-range order and spontaneous $U(1)$ symmetry breaking \cite{Lobos2009,Yan_2018,Radzihovsky_2023,Ribeiro2024,Kuklov2024iQTF,Kuklov_2024,Radzihovsky_2024}. In the latter case, the behavior depends on the system’s commensurability. For commensurate fillings, dissipation just creates a Mott-insulating phase \cite{Malatsetxebarria2013,Cai2014,Bouverot_Dupuis_2024,Bouverot_Dupuis_2025}, while for incommensurate fillings it stabilizes a density wave that, although gapless, exhibits long-range order \cite{Cazalilla2006,Majumdar_2023,Majumdar2023a}. These results have been obtained numerically via Monte Carlo simulations of microscopic Hamiltonians, and analytically using bosonization~\cite{Malatsetxebarria2013,Bouverot_Dupuis_2024,Cazalilla2006,Majumdar2023a}. The latter serves a starting point for perturbative renormalisation group analyses limited to weak dissipation and self-consistent harmonic approximations (SCHA) in the strong dissipation regime. However, a unified, nonperturbative description capable of capturing both regimes at once has so far been lacking.

In this work, we investigate the phase diagram of an incommensurate one-dimensional Bose–Hubbard model with local baths coupled to the density, using the nonperturbative functional renormalization group (FRG)~\cite{dupuis2021,delamotte_2012}. The FRG provides a coherent and systematically improvable analytical framework that continuously connects the weak- and strong-coupling regimes.
In a nutshell, it is a modern implementation of Wilson's renormalization group which gradually integrates out short-scale fluctuations to isolate the emergent large-scale physics. This approach has already yielded remarkably accurate descriptions of one-dimensional quantum systems \cite{Daviet_2019,jentsch2022}, long-range interacting systems \cite{Dupuis_2024,Daviet_2020,Balog_2014,pagni2025,Defenu_2020,Defenu_2017,Defenu_2015} and, more recently, of single-particle dissipative systems \cite{daviet2023,paris3CK_2025}.

Our results present the complete phase diagram of the model for sub-ohmic (slow bath dynamics), ohmic, and super-ohmic (fast bath dynamics) dissipation. In all cases, we identify in the renormalization group (RG) flow a continuous line of Luttinger-liquid (LL) fixed points and a dissipative fixed point (DFP) separated by a critical fixed point describing a Berezinskii--Kosterlitz--Thouless (BKT) transition whose location is bath-dependent. There are therefore two phases: the LL and the DFP. We further characterize the universal properties of the DFP and identify sub-leading corrections.

The remainder of this paper is organized as follows. Section~\ref{sec:model} introduces the dissipative Bose–Hubbard model and derives the effective bosonic action obtained by integrating out the bath. Section~\ref{sec:method} details the FRG formalism and its application to the present problem. The resulting phase diagrams for different bath types are presented and analyzed in Sec.~\ref{sec:results}. Finally, Sec.~\ref{sec:conclusion} summarizes our findings and discusses potential extensions of this work.

\section{Model}
\label{sec:model}
\subsection{Microscopic model}
\begin{figure}[h!]
    \centering
    \includegraphics[width=0.9\linewidth]{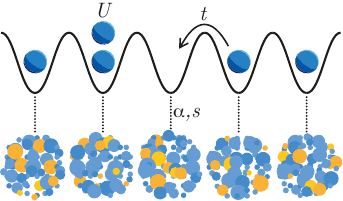}
    \caption{Schematic picture of the dissipative bosonic system considered in this work. Bosons are constrained to live on a one-dimensional lattice with a nearest-neighbour hopping amplitude $t$ and an on-site repulsion $U$. Each site is coupled to an independent bath characterized by a coupling $\alpha$ and a spectral exponent $s$, which characterize the low-energy spectrum of the bath.
    }
    \label{fig:dissipative_bosons}
\end{figure}

We consider a one-dimensional lattice of interacting bosons coupled locally to independent dissipative baths (see Fig.~\ref{fig:dissipative_bosons}). The Hamiltonian for the bosons $\{\hpsi_j,\hpsi_j^\dagger\}$ is
\begin{equation}
    \hat{H}_{\rm S}=\sum_j -t (\hpsi^\dagger_{j+1}\hpsi_j + \hpsi^\dagger_j \hpsi_{j+1}) + \frac{U}{2} \hpsi^\dagger_j \hpsi^\dagger_j \hpsi_j \hpsi_j,
\end{equation}
where $t$ and $U$ denote the hopping amplitude and on-site interaction, respectively, and the lattice spacing is set to unity. At each site $j$, we introduce a bath composed of harmonic oscillators $\{ \hat{P}_{j\gamma},\hat{X}_{j\gamma}\}$ \emph{à la} Caldeira and Leggett \cite{Caldeira_1983,Leggett_1987,Weiss_2012} with Hamiltonian
\begin{equation}
    \hat{H}_{\rm B}=\sum_{j,\gamma} \frac{\hat{P}_{j\gamma}^2}{2m_\gamma}+ \frac{1}{2}m_\gamma \Omega_\gamma^2\hat{X}_{j\gamma}^2,
\end{equation}
and coupled locally to the boson density via
\begin{equation}
    \hat{H}_{\rm SB}=\sum_j \hpsi^\dagger_j \hpsi_j \sum_\gamma \lambda_\gamma \hat{X}_{j\gamma}.
\end{equation}
The entire system is thus described by the Hamiltonian $\hat{H}=\hat{H}_{\rm S}+\hat{H}_{\rm SB} + \hat{H}_{\rm B}$. The microscopic couplings of the bath, namely $\{m_\gamma,\Omega_\gamma,\lambda_\gamma\}$, are made site-independent to preserve the translational symmetry of the system. Note that other system-bath couplings are possible. This includes coupling to the boson operators $\hpsi_j^\dagger$ and $\hpsi_j$ instead of the density \cite{Lobos2009,Ribeiro2024}, or connecting multiple sites to the same bath \cite{Min2024}.

The canonical partition function $Z=\operatorname{Tr} e^{-\beta \hat{H}}$ at inverse temperature $\beta=1/T$ can be represented using the path integral formalism. The bath degrees of freedom $\{\hat{P}_{j\gamma},\hat{X}_{j\gamma}\}$ being quadratic, they can be exactly integrated out to yield
\begin{equation}
    Z=\int \mathcal{D}\psi_j^\ast \mathcal{D}\psi_j \,e^{-S[\psi_j^\ast,\psi_j]},    
\end{equation}
where $\psi_j^\ast(\tau)$, $\psi_j(\tau)$ are complex fields periodic in the imaginary-time argument $\tau\in [0,\beta]$. The effective Euclidean action $S$ for the bosons reads
\begin{align}\label{eq:action_eff}
    S&=\sum_j \Bigg\{ \int_\tau \Big[\psi_j^\ast \partial_\tau \psi_j -t (\psi^\ast_{j+1}\psi_j + \psi_{j+1}\psi^\ast_j)\nonumber\\
    &+ \frac{U}{2} |\psi_j|^4\Big] -\int_{\tau,\tau'}|\psi_j|^2 \mathcal{D}(\tau-\tau')|\psi_j'|^2\Bigg\},
\end{align}
where $\int_\tau \equiv \int_0^\beta \d \tau$, $\psi_j\equiv \psi_j(\tau)$, $\psi_j'\equiv \psi_j (\tau')$ and the dissipative kernel $\mathcal{D}(\tau)$ is customarily defined in terms of the bath spectral function 
\begin{align}
   J(\Omega) = \frac{\pi}{2} \sum_\gamma \frac{\lambda_\gamma^2}{m_\gamma \Omega_\gamma} \delta(\Omega-\Omega_\gamma),
\end{align}
through
\begin{align}\label{eq:def_D}
    \mathcal{D}(\tau) = \int_0^\infty \frac{\d \Omega}{2\pi} J(\Omega) e^{-\Omega |\tau|}.
\end{align}
We assume that the spectral function exhibits the generic low-frequency behavior
\begin{align}
    J(\Omega) \propto |\Omega|^s \quad \text{for } \Omega \ll \Omega_c,
\end{align}
where $\Omega_c$ denotes a high-frequency cutoff scale set by the bath. For frequencies $\Omega \gtrsim \Omega_c$, $J(\Omega)$ is suppressed. Inserting this form into Eq.~\eqref{eq:def_D} directly yields
\begin{align}
    \mathcal{D}(\tau) \sim \frac{1}{|\tau|^{1+s}}
    \quad \text{for } |\tau| \gg \Omega_c^{-1}.
\end{align}
The short-time ($|\tau| \ll \Omega_c^{-1}$) behavior depends on the detailed high-frequency structure of $J(\Omega)$. It does not affect the universal low-energy properties of the system and therefore need not be specified explicitly. Physically, $\mathcal{D}(\tau)$ encodes the non-Markovian memory effects induced by the bath. The exponent $s$ distinguishes sub-ohmic ($s<1$), ohmic ($s=1$), and super-ohmic ($s>1$) dissipation. In the frequency domain, the low-frequency behavior corresponds to  $\mathcal{D}(i\omega)=-\alpha |\omega|^s/8$, where $\omega$ denotes a bosonic Matsubara frequency, $\alpha$ is the dissipative coupling, and the numerical prefactor is introduced for convenience.

\subsection{Bosonization}
Bosons can be described by the density-phase representation
\begin{equation}\label{eq:density_phase_repr}
    \psi_j(\tau)=\sqrt{\rho_j(\tau)}e^{i\theta_j(\tau)}.    
\end{equation}
In the continuum limit $\rho_j(\tau) \to \rho(x,\tau)$ and $\theta_j(\tau) \to \theta(x,\tau)$. Substituting \eqref{eq:density_phase_repr} into the action \eqref{eq:action_eff} and retaining only the most relevant terms leads to
\begin{align}
    S=&\int_{x,\tau} \Big[i\rho \partial_\tau \theta + t \rho (\partial_x \theta)^2 + \frac{U}{2} \rho^2\nonumber\\
    & -\int_{\tau'} \rho \,\mathcal{D}(\tau-\tau')\rho'\Big],
\end{align}
where $\rho' \equiv \rho(x,\tau')$.
In one dimension, the density fluctuations around the mean value $\rho_0$ can be represented by an additional bosonic field $\varphi(x,\tau)$ \cite{Haldane_1981,Giamarchi_2004,Cazalilla_2011,Dupuis_2026} as
\begin{equation}
    \rho = \left(\rho_0-\frac{1}{\pi}\partial_x \varphi\right)\sum_{m} A_m e^{i2m(\pi \rho_0 x -\varphi)},
\end{equation}
where $A_0=1$ and the other amplitudes $A_m$ are non-universal (i.e. model-specific). Away from any commensurate fillings ($\rho_0 \notin \mathbb{Q}$), keeping again only the most relevant terms yields the bosonized action
\begin{align}
    S=&\int_{x,\tau} \Big[-\frac{i}{\pi}\partial_x \varphi \partial_\tau \theta + t \rho_0 (\partial_x \theta)^2 + \frac{U}{2 \pi^2} (\partial_x \varphi)^2\nonumber\\
    & -\int_{\tau'} 2\rho_0^2 A_1 A_{-1} \cos(2\varphi-2\varphi') \mathcal{D}(\tau-\tau')\Big].
\end{align}
Finally, the field $\theta$ is integrated out to give
\begin{align}\label{eq:action}
    S=&\int_{x,\tau} \frac{1}{2\pi K}\Big[v (\partial_x \varphi)^2+ (\partial_\tau \varphi)^2/v \Big]\nonumber\\
    & -\int_{x,\tau,\tau'} \cos(2\varphi-2\varphi') \mathcal{D}(\tau-\tau'),
\end{align}
where we introduced the Luttinger parameter $K=\pi \sqrt{2t\rho_0/U}$, the velocity $v=\sqrt{2t \rho_0 U}$ and absorbed $2\rho_0^2 A_1 A_{-1}$ into the definition of $\alpha$. This bosonization procedure is expected to be reliable at large distances compared to the lattice spacing. However, the relations obtained between the microscopic parameters and $K$, $v$, $\alpha$ might be very crude. The resulting action~\eqref{eq:action} has appeared previously in related contexts, such as dissipative XXZ spin chains~\cite{Majumdar2023a,Majumdar_2023} and one-dimensional electronic systems coupled to metallic gates~\cite{Cazalilla2006}. This comes as no surprise since particle statistics play very little role in one dimension. Indeed, exchanging two interacting particles is only possible through a collision whose phase shift cannot be separated from the statistical phase. Upon exchanging space and imaginary-time, Eq.~\eqref{eq:action} also describes quantum  systems with long-range spatial interactions \cite{Orignac_2025}.

\subsection{Physical insights}

From the field theory defined in \eqref{eq:action}, we expect two phases depending on whether $\alpha$ is relevant or not. In the limit $\alpha \to 0$, the dissipation becomes negligible and one recovers the standard Luttinger-liquid (LL) action,
\begin{align}\label{eq:action_LL}
    S_{\rm LL}=&\int_{x,\tau} \frac{1}{2\pi K}\Big[v (\partial_x \varphi)^2+ (\partial_\tau \varphi)^2/v \Big],
\end{align}
which describes a phase with quasi–long-range order. The compressibility $\kappa=K/(\pi v)$ and the superfluid stiffness $\rho_s = K v/\pi$ are both finite. Note that throughout this work, the term “finite” is used to mean finite and nonzero. In the opposite limit of strong dissipation (large $\alpha$), the cosine interaction locks the field $\varphi$ into one of its minima, $\varphi = \text{const}$, so that $\cos(2\varphi-2\varphi')\simeq 1- 2(\varphi-\varphi')^2$. This leads to the action
\begin{align}\label{eq:action_iTQF}
    S_{\rm DFP}=&\int_{x,\tau} \frac{1}{2\pi K} \left[v(\partial_x \varphi)^2 + (\partial_\tau \varphi)^2/v \right] \nonumber\\
    &+ \int_{x,\tau,\tau'} 2(\varphi-\varphi')^2 \mathcal{D}(\tau-\tau')\nonumber\\
    \simeq &\int_{q,\omega} \left[\frac{v}{2\pi K} q^2 + \frac{\alpha}{2} |\omega|^s\right] |\varphi(q,i\omega)|^2,
\end{align}
where we neglected the regular kinetic term $\propto \omega^2$ which is dominated by the kernel $\mathcal{D}(i\omega)\sim|\omega|^s$. Despite being gapless, this phase exhibits long-range order under the form of a density wave of wavevector $k=2\pi\rho_0$, as the correlation function $\langle e^{i(2\varphi(x,\tau)-2\varphi(0,0))}\rangle$ does not vanish at large separation \cite{Majumdar2023a}. Its compressibility remains identical to that of the LL, $\kappa = K/(\pi v)$, but the superfluid stiffness vanishes. Performing a Wick rotation $\tau \to i t$, the real-time dispersion follows $\omega^s=(- i)^s\frac{v}{\pi K \alpha} q^2$. For ohmic dissipation ($s=1$), this corresponds to diffusive dynamics. When the density fluctuations $\varphi$ are replaced by the superfluid phase $\theta$, the action \eqref{eq:action_iTQF} is known as an \emph{incoherent Transverse Quantum Fluid} (iTQF) and describes a stable superfluid phase with off-diagonal long-range order \cite{Kuklov2024iQTF,Kuklov_2024,Radzihovsky_2023,Radzihovsky_2024}. Contrary to the dissipative phase \eqref{eq:action_iTQF}, it has a finite superfluid stiffness but a diverging compressibility.

Note that looking at the action \eqref{eq:action}, it might seem counterintuitive that small values of $K$ correspond to the DFP, and large ones to the LL phase. Considering $K$ is large, it appears \emph{a priori} that the dissipative part of the action provides the most important contribution. However, the previous argument, equivalent to the mean-field approximation, is incorrect. Indeed, the Luttinger parameter $K$ controls the cost in action to go from one minima of $\cos(2\varphi-2\varphi')$ to another. For large $K$, this cost is small so the field $\varphi$ can easily escape the potential, thus rendering dissipation unimportant at large scales.

\section{Method\label{sec:method}}
In this section, we briefly review the nonperturbative functional renormalization group (NPFRG)~\cite{dupuis2021,delamotte_2012}, also known as the functional renormalization group (FRG), before applying it to the model defined in Eq.~\eqref{eq:action}.

\subsection{FRG formalism}
The FRG provides a systematic framework to investigate the large-scale behavior of many-body systems by progressively integrating out short-scale fluctuations, in the spirit of Wilson’s momentum-shell RG or Kadanoff’s block-spin construction. To achieve this controlled coarse-graining, we introduce a running infrared (IR) regulator $R_k(q,i\omega)$ that suppresses modes with momenta and frequencies below a momentum scale $k$. The original action is thus replaced by
$S+\Delta S_k$ with
\begin{equation}
    \Delta S_k[\varphi] = \frac{1}{2} \int_{q,\omega} R_k(q,i\omega) |\varphi(q,i\omega)|^2,
\end{equation}
where $\int_{q,\omega} \equiv \int_{-\infty}^\infty\int_{-\infty}^\infty \,\frac{dq d\omega}{4\pi^2}$.
The regulator must satisfy the usual conditions
\begin{itemize}
    \item $R_k(q,i\omega) \sim k^2$ for $\sqrt{q^2+(\omega/v_k)^2}\ll k$ where $v_k$ is a scale-dependent velocity to be defined below,
    \item $R_k(q,i\omega) \to 0$ for $\sqrt{q^2+(\omega/v_k)^2}\gg k$. 
\end{itemize}
One then defines the scale-dependent partition function
\begin{equation}
    Z_k[J] = \int \mathcal{D}\varphi \, e^{-S[\varphi] - \Delta S_k[\varphi] + \int_{x,\tau} J\varphi},
\end{equation}
where $J\equiv J(x,\tau)$ is an external source field. Denoting by $\phi(x,\tau)=\langle \varphi(x,\tau) \rangle_J$ the average field in the presence of $J$, the scale-dependent effective action
\begin{equation}\label{eq:effective_action}
    \Gamma_k[\phi] = - \ln Z_k[J] + \int_{x,\tau} J \phi - \Delta S_k[\phi]
\end{equation}
is defined as a modified Legendre transform, which includes the subtraction of $\Delta S_k[\phi]$.
When $k =\Lambda$ with $\Lambda$ a UV cutoff much larger than all physical scales, $R_{k=\Lambda}$ suppresses all fluctuations so that $\Gamma_{k=\Lambda}[\phi] = S[\phi]$. For $k=0$, $R_k=0$ and $\Gamma_{k=0}[\phi]$ becomes the full effective action of the original model~\eqref{eq:action}.
The interpolation between the initial solvable limit ($k=\Lambda$) and the final physical one ($k = 0$) is described by the Wetterich equation \cite{Wetterich1993}
\begin{equation} \label{eq:Wetterich}
    \partial_t \Gamma_k[\phi] = \frac{1}{2} \int_{q,\omega}\hspace{-0.1cm} \partial_t R_k(q,i\omega) (\Gamma_k^{(2)}[\phi]+R_k)^{-1}(q,i\omega),
\end{equation}
where $\Gamma_k^{(2)}$ denotes the second functional derivative of $\Gamma_k$ with respect to $\phi$ and $t=\ln(k/\Lambda)$ is the (negative) RG time. Equation~\eqref{eq:Wetterich} lies at the core of the FRG approach, enabling nonperturbative approximations that capture both critical and strong-coupling regimes beyond standard perturbation theory.

\subsection{Derivative expansion}
A common approximation scheme to the Wetterich equation \eqref{eq:Wetterich} is the Derivative Expansion (DE), which consists in expanding the scale-dependent effective action $\Gamma_k[\phi]$ in powers of momenta and frequencies. In the present work, we employ a second-order derivative expansion (DE2) truncation, which retains terms up to quadratic order in derivatives,
\begin{align}\label{eq:DE2_ansatz}
    \Gamma_k[\phi]=&\int_{x,\tau} \frac{Z_{x,k}(\phi)}{2}\left(\partial_x \phi \right)^2 +\frac{Z_{\tau,k}(\phi)}{2}\left( \partial_\tau \phi\right)^2\nonumber\\
    &+ \sum_{i}\int_{x,\tau,\tau'} U_{i,k}(\phi,\phi')\mathcal{D}_{s_i}(\tau-\tau')\nonumber\\
    & +\int_{x,\tau} V_k(\phi).
\end{align}
The kernels are defined as $\mathcal{D}_{s_i}(i\omega) \equiv |\omega|^{s_i}$, such that $\mathcal{D}(i\omega) = -\alpha\, \mathcal{D}_s(i\omega)/8$. 
The exponents $s_i \in [s, 2[$ cover all intermediate frequency dependencies between the dissipative kernel $\mathcal{D}_s$ and the LL term proportional to $\omega^2$. Including dissipative kernels with exponent $s_i>s$ is necessary to ensure the consistency of the RG equations (see Appendix~\ref{sec:FlowEq}).
The structure of the ansatz is further constrained by the quasi-invariance of $\Gamma_k[\phi]$ under the Statistical Tilt Symmetry, $\phi(x,\tau)\to\phi(x,\tau)+w(x)$ (see Appendix~\ref{sec:STS}). It implies that $V_k(\phi) = V_k$ and $Z_{\tau,k}(\phi) = Z_{\tau,k}$ are independent of $\phi$, $U_{i,k}(\phi,\phi')=U_{i,k}(\phi-\phi')$, and $Z_{x,k}(\phi) = Z_x$ is not renormalized at all. The non-local potential must also be $\pi$-periodic and even in its argument $\phi - \phi'$ and thus admits the decomposition
\begin{equation}\label{eq:U_harmonics}
    U_{i,k}(\phi-\phi')=\sum_n u_{i,k}^{(n)}\cos(2n(\phi-\phi')).
\end{equation}
Functional flow equations are generally obtained by evaluating $\Gamma_k[\phi]$ in a uniform time-independent field configuration. However, this procedure only gives an equation for $\sum_n u_{i,k}^{(n)}$ instead of that for the full function $U_{i,k}$. The flow of the harmonics $u_{i,k}^{(n)}$, $n=1,\cdots, N$ can nevertheless be found by truncating \eqref{eq:U_harmonics} at $N$ terms and considering the coupled Wetterich equations for the vertices $\Gamma_k^{(2n)}$, $n=1,\cdots, N$. Hereafter, we decide for the sake of simplicity to only keep track of the flow of $u_{i,k}^{(0)}$ and $u_{i,k}^{(1)} \equiv \alpha_{i,k}/8$. The resulting ansatz reads
\begin{align}\label{eq:ansatz}
    \Gamma_k[\phi]&=\int_{x,\tau} \frac{Z_x}{2}\left(\partial_x \phi \right)^2 +\frac{Z_{\tau,k}}{2}\left( \partial_\tau \phi\right)^2 \nonumber\\
    &+ \sum_{i}\int_{x,\tau,\tau'} \frac{\alpha_{i,k}}{8} \cos(2\phi-2\phi') \mathcal{D}_{s_i}(\tau-\tau'),
\end{align}
where the unimportant free-energy term $V_k$ is omitted and the coefficients $u_{i,k}^{(0)}$ do not appear since they multiply $\int_{\tau}\mathcal{D}_{s_i}(\tau)=\mathcal{D}_{s_i}(i\omega=0)=0$.
Flow equations for $Z_{\tau,k}$ and the couplings $\alpha_{i,k}$ are obtained by isolating the coefficients of powers of $|\omega|$ in the flow equation of $\Gamma_k^{(2)}(q,i\omega)$. Details of this procedure and the explicit results for arbitrary $s$ are presented in Appendix~\ref{sec:FlowEq}. This single-mode truncation can of course be systematically improved by including higher harmonics $u_{i,k}^{(n)}$. Nevertheless, we find that the single-mode truncation already provides physically consistent results, as demonstrated below.

\subsection{Dimensionless quantities\label{sec:dimless}}

In order to observe the scale-invariance of a critical RG fixed point, it is essential to consider dimensionless quantities \cite{dupuis2021}. We define the dimensionless momentum and frequency
\begin{equation}
    \tilde q \equiv q/k, \qquad \tilde \omega \equiv \omega/(k v_k),
\end{equation}
couplings
\begin{equation}\label{eq:dimless}
    \tilde y_{\tau,k} \equiv \sqrt{\frac{Z_{\tau,k}}{Z_x}} v_k, \quad \tilde y_{i,k} \equiv \sqrt{\frac{\alpha_{i,k} k^{s_i-2}}{Z_x}}v_k^{s_i/2},
\end{equation}
and two-point vertex
\begin{equation}
    \tilde \Gamma_{k}^{(2)}(\tilde q,i\tilde\omega) \equiv \frac{\Gamma_{k}^{(2)}(q,i\omega)}{Z_x k^2},
\end{equation}
which depend on the running speed
\begin{equation}
    v_k \equiv \frac{1}{\sqrt{\frac{Z_{\tau,k}}{Z_x}} + \sum_i \left(\frac{\alpha_{i,k} k^{s_i-2}}{Z_x}\right)^{1/s_i}}.
\end{equation}
This last definition enforces the relation $\tilde y_{\tau,k} + \sum_i (\tilde y_{i,k})^{2/s_i} = 1$. This means that, whenever one of the couplings $Z_{\tau,k}$ or $\alpha_{i,k}$ dominates, the corresponding dimensionless coupling saturates to  $\tilde y_{\tau,k}\to 1$ or $\tilde y_{i,k} \to 1$, and the speed reduces accordingly to $v_k \to \sqrt{Z_x/Z_{\tau,k}}$ or $v_k \to (Z_x/\alpha_{i,k} k^{s_i-2} )^{1/s_i}$. The fact that $v_k$ interpolates between the Luttinger Liquid sound velocity and an appropriate notion of renormalized speed for the dissipative interactions allows observing both the Luttinger liquid
\begin{equation}
    \tilde \Gamma_{k\to 0}^{(2)}(\tilde q,i\tilde\omega) = \tilde q^2 + \tilde \omega^2,
\end{equation}
and the dissipative phase
\begin{equation}
    \tilde \Gamma_{k\to 0}^{(2)}(\tilde q,i\tilde\omega) = \tilde q^2 + |\tilde \omega|^s,
\end{equation}
on a single RG flow diagram.

\section{Results\label{sec:results}}

\subsection{Ohmic bath}

\begin{figure}[h!]
    \centering
    \includegraphics[width=\linewidth]{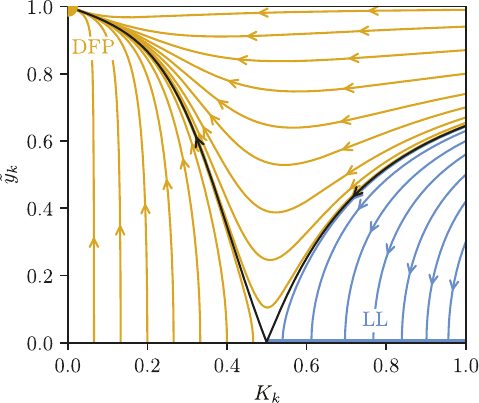}
    \caption{RG flow for ohmic ($s=1$) dissipation. The trajectories flow either toward the continuum of LL fixed points (blue) or toward the DFP (yellow). The flow around $(K_c,\tilde y_k)=(1/2,0)$ is that of a BKT transition with two (incoming and outgoing) separatrices drawn in black. Near the DFP, all trajectories flow into a single ``large river"~\cite{Bagnuls_2001a,Bagnuls_2001b}.}
    \label{fig:RG_flow_s1}
\end{figure}

We first focus on the case of ohmic dissipation ($s=1$), which has been the subject of extensive study~\cite{Weber_2022,Radzihovsky_2024,Ribeiro2024,Cazalilla2006,Majumdar2023a,Cai2014,Malatsetxebarria2013}. In this case, the RG flow does not generate any intermediate-frequency kernel, so the ansatz~\eqref{eq:ansatz} involves a single dissipative coupling, $\alpha_k \equiv \alpha_{0,k}$. Figure~\ref{fig:RG_flow_s1} shows the renormalization group trajectories in terms of the dimensionless coupling $\tilde y_k \equiv \tilde y_{0,k}$ and the running Luttinger parameter $K_k \equiv 1/(\pi\sqrt{Z_x Z_{\tau,k}})$. Two distinct phases can be clearly identified: a continuum of LL fixed points characterized by $K_{k=0} > 1/2$, and $\tilde y_{k=0} = 0$, and the dissipative phase with $K_{k=0} = 0$ and $\tilde y_{k=0} = 1$.

The nature and location of the transition can be inferred from a perturbative expansion of the dimensionless flow equations near the critical point. Expanding in the small parameters $x_k = K_k - 1/2$ and $\tilde y_k$, the RG equations reduce to
\begin{align}
&\partial_t \tilde y_k = x_k \tilde y_k,\
&\partial_t x_k = \frac{\bar C}{2} \tilde y_k^2,
\end{align}
where $\bar C$ is a non-universal constant (see Appendix~\ref{app:LL}). These equations are that of a Berezinskii--Kosterlitz--Thouless (BKT) transition at $K_c=1/2$, which is consistent with the perturbative RG calculation of Ref.~\cite{Majumdar_2023}. 

\begin{figure}[h!]
    \centering
    \includegraphics[width=\linewidth]{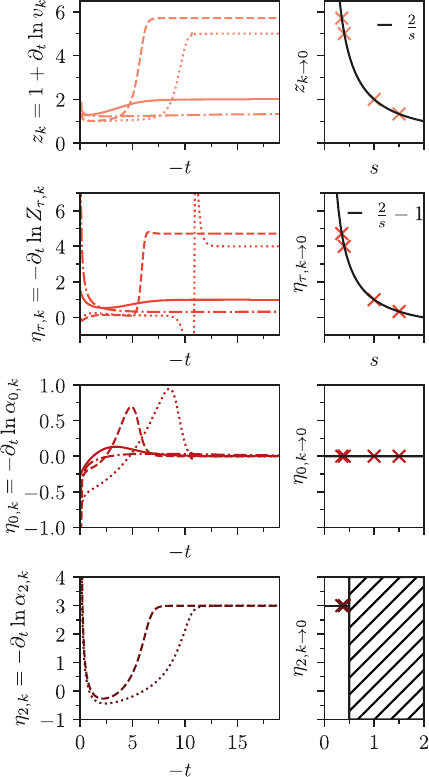}
    \caption{Left panels (top to bottom): RG flow of the running dynamical exponent $z_k$ and exponents $\eta_{\tau,k}$, $\eta_{0,k}$, and $\eta_{2,k}$ for trajectories flowing to the DFP. The RG time is $-t=\ln (\Lambda/k)$. The different lines correspond to $s=0.35$ (dashed), $s=0.4$ (dotted), $s=1$ (solid), and $s=1.5$ (dash-dotted). The rapid evolution of $\eta_{\tau,k}$ for $s=0.4$ at $-t \simeq 11$ is continuous; it gets more peaked as $s \to 0.5^-$. Right panels: asymptotic values of $z_k$, $\eta_{\tau,k}$, $\eta_{0,k}$, and $\eta_{2,k}$ as a function of $s$. The black curves are the analytical predictions obtained from the expansion about the DFP described in the main text.}
    \label{fig:RG_anomalous_dimension}
\end{figure}

A perturbative expansion of the flow equations around the dissipative regime, i.e., for $\tilde y_k \to 1$ and $K_k \to 0$, yields analytical predictions for the critical behavior (see Appendix~\ref{app:TQF}). In this limit, the flow equations give
\begin{equation}
    v_k \sim k^{z-1}, \qquad Z_{\tau,k} \sim k^{-\eta_\tau},    
\end{equation}
with the dynamical critical exponent $z = 2$ and the scaling exponent $\eta_\tau = 1$, while the dissipative coupling $\alpha_{0,k}$ approaches a finite value. These analytical results are fully consistent with the numerical solution of the flow equations (see Fig.~\ref{fig:RG_anomalous_dimension} where the solid lines correspond to $s=1$).

The non-zero scaling exponent $\eta_\tau$ modifies the low-frequency scaling of the kinetic term $Z_{\tau,k}\omega^2$ in the dissipative regime. Although the derivative expansion is only valid in the regime $|q|\ll k$ and $|\omega|\ll kv_k$, the low-energy behavior in the limit $k\to 0$
can be captured by evaluating the effective frequency at the running scale $\omega_k = k v_k$~\cite{blaizot2006,dupuis2021}. At that scale,
\begin{equation}\label{eq:gluing_DE}
    Z_{\tau,k} \omega_k^2 \sim k^{2z-\eta_\tau}\sim \omega_k^{2-\eta_\tau/z},
\end{equation}
so that, with $z=2$ and $\eta_\tau=1$, the quadratic term $\sim \omega^2$ is effectively replaced by a non-analytic term $\sim |\omega|^{3/2}$. The two-point vertex at the DFP is therefore
\begin{equation}
    \Gamma_{k=0}^{(2)}(q, i\omega) = Z_x q^2 + A^\ast |\omega| + B^\ast|\omega|^\frac{3}{2},
\end{equation}
where $A^\ast$ and $B^\ast$ are constants characterizing the DFP. This sub-leading behavior is in agreement with previous self-consistent harmonic-approximation results~\cite{Majumdar_2023}.

\subsection{Super-ohmic bath}
\begin{figure}[h!]
    \centering
    \includegraphics[width=\linewidth]{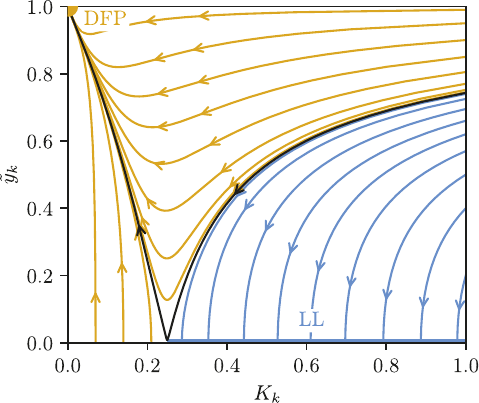}
    \caption{RG flow for super-ohmic ($s=1.5$) dissipation. The RG trajectories either flow to a continuum of LL fixed points in blue, or to the DFP in yellow. The BKT transition is at $K_c=1-s/2$ and $\tilde y_k=0$.}
    \label{fig:RG_flow_s1.5}
\end{figure}

The case of a super-ohmic bath ($s>1$), exhibits a behavior qualitatively similar to the ohmic case (see Fig.~\ref{fig:RG_flow_s1.5}). Two distinct phases are again observed: a line of LL fixed points with $\tilde y_{k=0} = 0$ and $K_{k=0} > K_c$, and an isolated fixed point at $\tilde y_{k=0} = 1$, $K_{k=0} = 0$ corresponding to the dissipative phase. A perturbative expansion near the critical point (see Appendix~\ref{app:LL}) shows that the BKT transition occurs at $K_c = 1 - s/2$, in agreement with Ref.~\cite{Majumdar_2023}. The main difference compared to the ohmic case thus lies in the enhanced robustness of the LL phase against dissipation.

A perturbative analysis of the flow equations around the dissipative regime yields
\begin{align} \label{eq:exponents}
z = \frac{2}{s}, \qquad
\eta_\tau = \frac{2}{s} - 1.
\end{align}
Following the argument outlined in Eq.~\eqref{eq:gluing_DE}, these exponents imply that the two-point vertex behaves as
\begin{equation}\label{eq:DFP_Gamma}
    \Gamma_{k=0}^{(2)}(q, i\omega) = Z_x q^2 + A_s^\ast |\omega|^s + B_s^\ast|\omega|^{1+\frac{s}{2}},
\end{equation}
at the DFP, where $A^\ast_s$ and $B_s^\ast$ denote $s$-dependent constants. Our analytical predictions are fully supported by the numerical integration of the flow equations, shown in Fig.~\ref{fig:RG_anomalous_dimension}.

\subsection{Sub-ohmic bath}
For sub-ohmic baths, the DFP action \eqref{eq:action_iTQF} describes a localized phase with vanishing DC conductivity \cite{Majumdar_2023}. Within our approach, this regime distinguishes itself by the emergence of additional couplings $\alpha_{i\geq1,k}$, which represent higher-order corrections to the spectral density $J(\Omega)$. For $s\in ]1/n,1/(n+1)[$ the powers $|\omega|^{s_i}$ generated are of the form $s_i = 1+2s, 1+3s, \ldots, 1+ns$ with $n$ a positive integer. Although an additional term with exponent $s_1 = 1+s$ could in principle appear, we find that it is not generated by the flow. At the special values $s=1/n$, with $n>1$, the coupling $\tilde y_{n,k}$ associated with $|\omega|^{1+ns}$ becomes marginal in the LL, leading to logarithmic corrections (see Appendix~\ref{App:1_n} for more details).

\begin{figure}[h!]
    \centering
    \includegraphics[width=\linewidth]{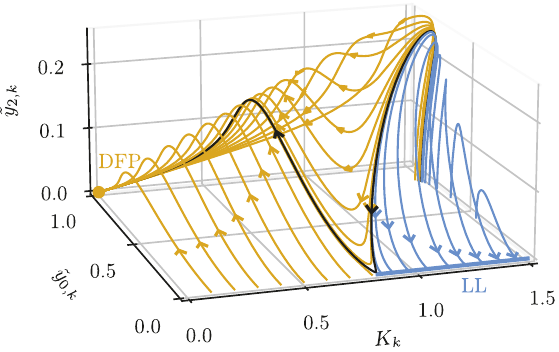}
    \caption{
    RG flow for sub-ohmic ($s=0.35$) dissipation in the three-dimensional space $(K_k, \tilde y_{0,k}, \tilde y_{2,k})$. After a transient regime where $\tilde y_{2,k}$ increases rapidly, the RG trajectories either flow to a continuum of LL fixed points in blue, or to the DFP in yellow. The BKT transition occurs at $K_c=1- s/2$ and $\tilde y_{0,k}=\tilde y_{2,k}=0$. The DFP attracts all trajectories through a “large river” flow. Because the LL fixed points form a continuum, they are reached in a "large surface" which, here, is asymptotically defined by $\tilde y_{2,k}=0$. The initial values $\tilde y_{2,\Lambda}$ are taken small but finite, since the numerical flow equations become ill-behaved at $\tilde y_{2,k}=0$.
    }
    \label{fig:RG_flow_s0.35}
\end{figure}

To illustrate the impact of the additional sub-ohmic couplings, we focus on the representative case $1/3 < s < 1/2$, where a single additional coupling $\tilde y_{2,k}$ is generated. Figure~\ref{fig:RG_flow_s0.35} shows the numerical integration of the flow equations in the three-dimensional space $(K_k,\tilde y_{0,k},\tilde y_{2,k})$. At early RG times, the flow exhibits a transient regime in which $\tilde y_{2,k}$ grows rapidly.
Trajectories then either flow toward the continuum of LL fixed points at $K>K_c = 1-s/2$ with $\tilde y_{0,k=0}=\tilde y_{2,k=0}=0$, or toward the isolated DFP at $K=\tilde y_{2,k=0}=0$ and $\tilde y_{0,k=0}=1$. The critical point, located at $K=K_c=1-s/2, \, \tilde y_{0,k} = \tilde y_{2,k} = 0$, comes with separating surfaces containing the two black lines on Fig. \ref{fig:RG_flow_s0.35}. The transition belongs to the BKT universality class (see Appendix~\ref{app:sub-ohmic_pert}). 

For all $s<1$, the expressions in Eq.~\eqref{eq:exponents} for $z$ and $\eta_\tau$ remain valid in the dissipative phase. In addition, one can analytically derive the scaling exponents of the higher-order couplings $\alpha_{i\geq2}$,
\begin{equation}
    \alpha_{i,k} \sim k^{-\eta_i}, \quad \eta_i = 2i-1, \quad i\geq2,
\end{equation}
which are found to be independent of $s$.
This is confirmed numerically in Fig.~\ref{fig:RG_anomalous_dimension} for $1/3 < s < 1/2$. Extending the argument presented for the ohmic and super-ohmic cases, we find that all additional couplings ultimately renormalize into a $|\omega|^{1+\frac{s}{2}}$ correction to the inverse propagator $\Gamma_k^{(2)}(q,i\omega)$, so that Eq.~\eqref{eq:DFP_Gamma} remains valid in the sub-ohmic regime.

\section{Conclusion\label{sec:conclusion}}

We investigated the phase diagram of a one-dimensional dissipative Bose–Hubbard model using the nonperturbative functional renormalization group. We focus on an incommensurate Bose-Hubbard model which, without dissipation, is described by a LL with parameter $K$. For a broad class of bath spectra, two distinct low-energy regimes emerge. Turning on the dissipation, the LL remains stable for $K>K_c$ with $K_c$ a bath-dependent threshold, and transitions into a dissipative fixed point otherwise.
Our FRG analysis provides a detailed understanding of this dissipative phase, characterized by a finite compressibility, a vanishing superfluid stiffness and a finite dissipative coupling. We also observe non-trivial scaling exponents governing the low-energy behavior of the DFP and extract their analytical expressions from the full non-perturbative flow. By performing a perturbative expansion around the critical region, we show explicitly that the LL line of fixed points and the DFP are separated by a BKT transition whose location is bath-dependent, in agreement with previous studies. These results generalize and complement recent work on incoherent transverse quantum fluids~\cite{Kuklov2024iQTF}.

To derive these results, we employed a minimal ansatz for the effective action containing a single field harmonic but fully accounting for the frequency dependency of the dissipative kernel. While the latter crucially enables the observation of the dissipative fixed point, this leaves open the possibility of systematically improving the method by including higher-order modes. This could allow to study the full functional dependency of the dissipative fixed point.

Overall, our work demonstrates that the FRG provides a robust nonperturbative framework for studying one-dimensional quantum many-body systems with local dissipative couplings. Looking ahead, the framework developed here can be extended to explore the effect of dissipation on various phases of quantum matter.

%Overall, our work establishes the FRG as a robust and versatile nonperturbative approach to quantum many-body systems coupled to dissipative environments. Looking ahead, the framework developed here can be extended to explore the effect of dissipation on various phases of quantum matter.
\hspace{3mm}
\section*{Acknowledgements}
We thank Laura Foini, Christophe Mora, Alberto Rosso and especially Nicolas Dupuis for fruitful discussions. O.B.-D. acknowledges the support of the French ANR under the grant ANR-22-CMAS-0001 (\emph{QuanTEdu-France} project). O.B.-D. also thanks LPTMC, where a significant part of this work was done, for its hospitality and enlightening seminars.

\newpage

\appendix
\begin{widetext}

\section{Statistical Tilt Symmetry} \label{sec:STS}

The Statistical Tilt Symmetry (STS) corresponds to the transformation $\varphi(x,\tau) \to \varphi'(x,\tau)=\varphi(x,\tau) +w(x)$ with $w(x)$ arbitrary. In this section, we explicitly demonstrate the (quasi-)invariance of the effective average action $\Gamma_k[\phi]$ under the STS, following the approach of Ref.~\cite{daviet_2020_bose}. We start from the scale-dependent partition function, rewritten in terms of $\varphi'$ and $w$:
\begin{align}
    Z_k[J] \equiv &\int \mathcal{D}\varphi \exp \Bigg\{-S[\varphi]-\Delta S_k[\varphi] +  \int_{x,\tau} J(x,\tau) \varphi(x,\tau) \bigg\}\nonumber\\
    =&\int \mathcal{D}\varphi' \exp \Bigg\{-S[\varphi']-\Delta S_k[\varphi'] +  \int_{x,\tau} J(x,\tau) \varphi'(x,\tau)\nonumber \\
    & -  \int_{x,\tau} J(x,\tau) w(x) + \frac{Z_x}{2} \int_{x,\tau} \big(2\partial_x \varphi'(x,\tau) \partial_x w(x) -    \left(\partial_x w(x)\right)^2\big)\nonumber \\
    &  + \frac{1}{2} \int_{x,\tau,x',\tau'} R_k(x-x',\tau-\tau') \big(2\varphi'(x,\tau) w(x') -w(x) w(x')\big)  \bigg\},
\end{align}
where we have used the invariance of the functional measure under the rigid shift $\varphi \to \varphi'=\varphi +w$. Next, we define a new source $J'$ by collecting all terms linearly coupled to $\varphi'$, i.e.
\begin{equation}\label{eq:def_newJ}
    J'(x,\tau)=J(x,\tau)-Z_x \partial^2_x w(x) + \int_{x',\tau'} R_k(x-x',\tau-\tau') w(x'),
\end{equation}
and re-express everything in terms of $J'$ to arrive at
\begin{equation}
    Z_k[J] = Z_k[J'] \exp{- \int_{x,\tau} J'(x,\tau) w(x) + \frac{Z_x}{2} \int_{x,\tau} \left(\partial_x w(x)\right)^2 + \frac{1}{2} \int_{x,\tau,x',\tau'} R_k(x-x',\tau-\tau') w(x) w(x')}
\end{equation}
Taking a functional derivative with respect to $J$ yields
\begin{align}\label{eq:def_newphi}
    \phi(x,\tau) \equiv \frac{\delta \log Z_k[J]}{\delta J(x,\tau)} = \frac{\delta \log Z_k[J']}{\delta J'(x,\tau)} - w(x) = \phi'(x,\tau) - w(x).
\end{align}
Using Eqs.~\eqref{eq:def_newJ}–\eqref{eq:def_newphi}, one readily verifies that the modified scale-dependent effective action $\Gamma_k[\phi]$ \eqref{eq:effective_action} can be written as
\begin{equation}\label{eq:STS}
    \Gamma_k[\phi] = \Gamma_k[\phi'] + \frac{Z_x}{2} \beta \int_x \left(\partial_x w(x)\right)^2 + Z_x \int_{x,\tau} \phi'(x,\tau) \partial_x^2 w(x).
\end{equation}
Considering the DE2 ansatz \eqref{eq:DE2_ansatz}, this first shows that $Z_{x,k}(\phi)$, $Z_{\tau,k}(\phi)$, and $V_k(\phi)$ are field-independent since $\Gamma_k[\phi]$ remains invariant under the uniform shift $\phi(x,\tau) \to \phi(x,\tau) + w$.
Second, because Eq.~\eqref{eq:STS} is valid for all renormalization scales $k$, it follows that $Z_{x,k}$ is not renormalized at all and that no higher-order spatial derivative terms are generated. Third, all non-local operators $U_k(\phi,\phi')$ take the form $U_k(\phi-\phi')$.

\section{Dimensionful flow equations} \label{sec:FlowEq}
In this section, we derive the FRG flow equations for the couplings $Z_{\tau,k}$ and  $\alpha_{i,k}$. Differentiating functionally twice the Wetterich equation~\eqref{eq:Wetterich} with respect to $\phi$, one gets a flow equation for the two-point vertex $\Gamma^{(2)}_k[q, i\omega; \phi]$. Evaluating it in a constant and uniform field configuration $\phi(x,\tau)=\phi$ and using the ansatz \eqref{eq:ansatz}, one obtains the flow equation
\begin{align}\label{eq:Gamma_2_flow}
    \partial_t Z_{\tau,k} \Omega^2+ \sum_i \partial_t \alpha_{i,k}|\Omega|^{s_i}=& \sum_i \frac{\alpha_{i,k}}{2} \int_{\omega}f(\omega)\Big(2|\omega|^{s_i} + 2|\Omega|^{s_i}-|\omega-\Omega|^{s_i} -|\omega+\Omega|^{s_i}\Big),
\end{align}
with 
\begin{align}\label{eq:f_omega}
    f(\omega) = 4\int_{q}\frac{\partial_tR_k(q,i\omega)}{(Z_x q^2+Z_{\tau,k}\omega^2+\sum_i \alpha_{i,k} |\omega|^{s_i} +R_k(q,i\omega))^2}.
\end{align}
The regulator $R_k$ has the standard form~\cite{dupuis2021}
\begin{align}
    R_k(q,i\omega)=Z_x k^2 y r(y), \text{ with } y=\frac{Z_x q^2 + Z_{\tau,k}\omega^2 + \sum_i \alpha_{i,k} |\omega|^{s_i}}{Z_x k^2},
\end{align}
with $r(x)=\gamma/(e^x-1)$. Numerically, we find that the parameter $\gamma$ can be varied between $1$ and $6$ without affecting qualitatively the results. 

Individual RG equations for each coupling are extracted as follows. The low-frequency behavior of the right-hand side of Eq.~\eqref{eq:Gamma_2_flow} is analyzed step by step from the dominant frequency behavior $|\Omega|^s$ down to $\Omega^2$, considering as many coefficients $\alpha_{i,k}$ as needed to match the frequency expansions of the left- and right-hand sides. Since the couplings $\alpha_{i\geq 1,k}$ are not present in the microscopic action~\eqref{eq:action}, they are taken into account only if they are generated by the renormalization flow. This procedure is illustrated in details below. We show that terms with exponents $s_i=1+i s$ for $i\geq 0$ integer are generated. As we wish to observe departures from the LL frequency-behaviour $\sim \omega^2$, we restrict ourselves to $s_i<2$ and Eq.~\eqref{eq:Gamma_2_flow} is to be understood as valid only for frequency terms $|\omega|^\eta$ with $\eta \leq 2$.

The low-frequency structure of $f(\omega)$ plays a central role in the determination of the flow equations. Since, as we will show, $s_{i\geq 1}\geq 1$, it reads:
\begin{align}\label{eq:f_low}
f(\omega)=f_0+ f_1 |\omega|^s + f_2 |\omega|^{2s}+\cdots+ O(\omega).
\end{align}
The coefficients $f_i$ are of the form $f_i=g_i+\frac{\partial_t\alpha_{0,k}}{\alpha_{0,k}}h_i$ where $g_i$ and $h_i$ are numerical coefficients that solely depend on $\alpha_{0,k}$ (and not its time derivative). To deal with $s>1/3$ as is done in the main text, it turns out that one only needs $f_0$ and $f_1$ which are
\begin{align}
    &f_0=-\frac{8}{Z_xk^2}\int_q\frac{r'\left(q^2/k^2\right)}{(1+r\left(q^2/k^2\right))^2},\\
    &f_1=-\frac{8\alpha_{0,k}}{Z_x^2 k^4}\int_q \frac{r''\left(q^2/k^2\right)(1+r\left(q^2/k^2\right))-2r'\left(q^2/k^2\right)^2}{\left(1+r\left(q^2/k^2\right)\right)^3}+\frac{4\partial_t\alpha_{0,k}}{Z_x^2}\int_q\frac{r\left(q^2/k^2\right)+q^2/k^2 r'\left(q^2/k^2\right)}{q^4(1+r\left(q^2/k^2\right))^2},
\end{align}
and an exact expression can be derived for all the coefficients $g_i$,
\begin{equation}
     g_{i} = \frac{4 (-1)^i\alpha_0^i k^{-1-2i}}{i! Z_x^{1+i}} \frac{\gamma}{\sqrt{\pi}(1-\gamma) }  {\rm Li}_{-\frac{1}{2}-i}(1-\gamma), 
\end{equation}
where $\rm Li$ is the polylogarithm.

\subsection{Super-ohmic bath}
The right-hand side of Eq.~\eqref{eq:Gamma_2_flow} is expanded at small $\Omega$ using
\begin{align}\label{eq:small_w_expansion}
    \int_{\omega}f(\omega)\Big(2|\omega|^{s_i} - |\omega-\Omega|^{s_i} - |\omega+\Omega|^{s_i}\Big) &= - \Omega^2  s_i(s_i - 1)\int_{\omega}f(\omega) |\omega|^{s_i - 2} + o(\Omega^2).
\end{align}
This yields the flow equations
\begin{align}\label{eq:flow_super-ohmic}
    \partial_t \alpha_{0,k}=&  \alpha_{0,k} \int_{\omega}f(\omega),\\
    \partial_t \alpha_{i,k}=&  \alpha_{i,k} \int_{\omega}f(\omega),\\
    \partial_t Z_{\tau,k}=&- \sum_i\alpha_{i,k} \frac{s_i(s_i-1)}{2} \int_{\omega}f(\omega) |\omega|^{s_i-2}.\label{eq:flow_super-ohmic_2}
\end{align}
As the coefficients $\alpha_{i\geq 1,k}$ are absent from the initial condition~\eqref{eq:action}, they are never generated by the renormalization flow. We are thus left with two coupled equations for $\alpha_{0,k}$ and $Z_{\tau,k}$,
\begin{align}
    \partial_t \alpha_{0,k}=&  \alpha_{0,k} \int_{\omega}f(\omega),\\
    \partial_t Z_{\tau,k}=&- \alpha_{0,k} \frac{s(s-1)}{2} \int_{\omega}f(\omega) |\omega|^{s-2}.
\end{align}

\subsection{Ohmic bath}
When $s=1$, the previous small-$\Omega$ expansion \eqref{eq:small_w_expansion} breaks down since $\int_\omega f(\omega)|\omega|^{s-2}$ diverges logarithmically in the infrared. Using the expansion \eqref{eq:f_low}, we instead show that
\begin{align}
    \frac{\alpha_{0,k}}{2}\int_{\omega}f(\omega) \Big(2|\omega| - |\omega-\Omega| - |\omega+\Omega| \Big)\nonumber
    &= \frac{\alpha_{0,k}}{2} \Omega^2\int_{-\infty}^{+\infty} \frac{\d x}{2\pi} f(x\Omega) \Big( 2|x| - |1-x| - |1+x| \Big)\nonumber\\
    &= - \frac{\alpha_{0,k}}{2\pi} \Omega^2 f_0 + o(\Omega^2).
\end{align}
Hence,
\begin{align}\label{eq:flow_ohmic_1}
    \partial_t \alpha_{0,k}=&  \alpha_{0,k} \int_{\omega}f(\omega),\\\label{eq:flow_ohmic_2}
    \partial_t Z_{\tau,k}=& - \frac{\alpha_{0,k}}{2\pi} f_0,
\end{align}
which can also be recovered from Eq.~\eqref{eq:flow_super-ohmic} by setting $s=1+\varepsilon$ and using $\varepsilon \omega^{\varepsilon -1} \overset{\varepsilon \ll 1}{\simeq} 2\delta(\omega)$. 

\subsection{Sub-ohmic bath}
Let us focus first on the case $1/2<s<1$. The integral $\int_\omega f(\omega)|\omega|^{s-2}$ does not converge so the expansion \eqref{eq:small_w_expansion} is not valid. This is because a $|\Omega|^{1+s}$ term is created when approximating $f(\omega)\simeq f_0$ as
\begin{align}\label{eq:s1_term}
    \frac{\alpha_{0,k}}{2} \int_{\omega}f(\omega) \Big(2|\omega|^s - |\omega-\Omega|^s - |\omega+\Omega|^s \Big)
    &\simeq \alpha_{0,k} |\Omega|^{1+s}f_0 I_{s,0}.
\end{align}
with $I_{s,0}=\frac{1}{2 \pi}\int_0^\infty \d x \big(2|x|^s - |1-x|^s- |1+x|^s\big)$. Once this contribution has been extracted, a small-$\Omega$ expansion shows that the rest generates a $\Omega^2$ term as
\begin{align}
    &\frac{\alpha_{0,k}}{2} \int_{\omega}f(\omega) \Big(2|\omega|^s - |\omega-\Omega|^s - |\omega+\Omega|^s \Big)-\alpha_{0,k} |\Omega|^{1+s}f_0 I_{s,0}\nonumber\\
    &= \frac{\alpha_{0,k}}{2} \int_{\omega}(f(\omega)-f_0) \Big(2|\omega|^s - |\omega-\Omega|^s - |\omega+\Omega|^s \Big)\nonumber\\
    &=-\Omega^2 s_i(s_i-1)\frac{\alpha_{0,k}}{2} \int_{\omega}(f(\omega)-f_0) |\omega|^{s-2} + o(\Omega^2).
\end{align}
We thus include a coupling $\alpha_{0,k}$ for $s_0=s$ and another one, $\alpha_{1,k}$, for $s_1=1+s$. Since $f(\omega)-f_0 \simeq f_1 |\omega|^s$ at small $\omega$, the prefactor of the $\Omega^2$ term is a convergent integral for $s>1/2$. Piecing everything together, the RG equations for $1/2<s<1$ are found to be
\begin{align}
    \partial_t \alpha_{0,k}=& \alpha_{0,k} \int_{\omega}f(\omega),\\
    \partial_t \alpha_{1,k}=& \alpha_{1,k} \int_{\omega}f(\omega)  + \alpha_{0,k} f_0 I_{s,0},\\
    \partial_t Z_{\tau,k}=& -\alpha_{0,k} \frac{s_0(s_0-1)}{2} \int_{\omega} (f(\omega)-f_0) |\omega|^{s_0-2} - \alpha_{1,k} \frac{s_1(s_1-1)}{2} \int_{\omega} f(\omega) |\omega|^{s_1-2}.
\end{align}

This procedure is straightforwardly generalized to $1/(n+1)<s<1/n$, with $n\in \mathbb{N}^*$. The extra couplings generated are $s_i=1+is$ with $i=1,\dots,n$ and the corresponding flow equations are
\begin{align}\label{eq:flow_sub-ohmic_1}
    \partial_t \alpha_{i,k}=& \alpha_{i,k} \int_{\omega}f(\omega)  + \alpha_{0,k} f_{i-1} I_{s,i-1} \text{ for } i \in \llbracket 0,n \rrbracket,\\
    \partial_t Z_{k,\tau} =& -\alpha_{0,k} \frac{s (s - 1)}{2} \int_{\omega} \left(f(\omega)-\sum_{i=0}^{n-1} f_i |\omega|^{s_i-1} \right) |\omega|^{s-2} -  \sum_{i=1}^n\alpha_{i,k} \frac{s_i(s_i-1)}{2} \int_{\omega} f(\omega) |\omega|^{s_i-2},\label{eq:flow_sub-ohmic_3}
\end{align}
where we have introduced, for $n\ge 0$, 
\begin{align}\label{eq:Isn}
    I_{s,n}&=\frac{1}{2 \pi}\int_0^\infty \d x |x|^{ns} \big(2|x|^s - |1-x|^s- |1+x|^s\big)\\
    &=\frac{\pi}{\left(1 + \cos(s\pi) + \cos(sn\pi) +\cos(s (n+1)\pi)\right)\Gamma(-s) \Gamma(-sn) \Gamma(2 + s(n+1))}
\end{align}
and defined $I_{s,-1}=f_{-1}=0$ for convenience. Evaluating Eq.~\eqref{eq:Isn} at $n=0$, we find that $I_{s,0}=0$. Together with Eq.~\eqref{eq:flow_sub-ohmic_1}, it shows that if $\alpha_{1,\Lambda}=0$ at the initial RG scale $\Lambda$, $\alpha_{1,k}$ is never generated by the RG flow, i.e. $\partial_t\alpha_{1,k}$ actually remains zero all along the flow. This is why in the main text we consider $1/3<s<1/2$ as the minimal example which displays an additional coupling.

\subsection{Case $s=1/n$} \label{App:1_n}
In the case $s=1/n$, $n\geq 2$, the exponent $1+ns$ generated by the dissipative term collides with the LL exponent $2$ and we expect logarithmic corrections to appear as $\Omega^{2+\varepsilon} - \Omega^2 \sim \varepsilon \Omega^2 \log \Omega$. In the RG terminology, the operator associated to the exponent $1+ns$ becomes marginal with respect to the LL fixed point and generates logarithmic corrections. We thus include it by considering the following ansatz
\begin{align}
    \Gamma_k^{(2)}(q,i\omega;\phi)&=Z_xq^2+Z_{\tau,k}\omega^2 + \sum_{i=1}^{n-1}\alpha_{i,k} |\omega|^{s_i} + \alpha_{n,k} |\omega|^2\log(|\omega|/\Omega_0),
\end{align}
where $\Omega_0>0$ is a constant. With this ansatz, the logarithmic term is generated in the RG equations from the integral
\begin{align}
    \frac{ \alpha_{0,k}}{2} \int_{\omega}(f(\omega)-\sum_{i=0}^{n-1}f_i|\omega|^{s_i}) \Big(2|\omega|^{s_0} - |\omega-\Omega|^{s_0}- |\omega+\Omega|^{s_0}\Big),
\end{align}
whose integrand, at small $\omega$, behaves as $f_n |\omega|^{-1}$. Formally, the logarithmic contribution is extracted by splitting the integration domain in three pieces that are treated independently: $[0,\Omega]$, $[\Omega,\Omega_0]$ and $[\Omega_0,\infty[$. We write below the generalized flow equations in this case for the sake of completeness, but were unable to numerically solve them because of severe numerical instability problems:
\begin{align}
    \label{eq:alphai_RGflow_sub-ohmic_spe}
    \partial_t \alpha_{i,k}=& \alpha_{i,k} \int_{\omega}f(\omega)  + \alpha_{0,k} f_{i-1} I_{s,i-1} \text{ for } i \in \llbracket 0,n-1 \rrbracket,\\
    \partial_t \alpha_{n,k} =& \alpha_{n,k} \int_{\omega}f(\omega) + \frac{s_0(s_0-1)}{2\pi} \alpha_{0,k}f_{n-1} \\
    \partial_t Z_{\tau,k}=&-\sum_{i=1}^{n-1} \alpha_{i,k} \frac{s_i(s_i-1)}{2} \int_{\omega} f(\omega) |\omega|^{s_i-2} - \alpha_{n,k} \int_{\omega}f(\omega) \Big(\log(|\omega|/\Omega_0)+\frac{3}{2}\Big) \nonumber\\
    & + \alpha_{0,k}\Bigg[  f_{n-1} (I^<_{s,n-1} + I^>_{s,n-1})
    -s_0(s_0-1) \int_{\Omega_0}^\infty \frac{\d \omega}{2 \pi}\big(f(\omega)-\sum_{l=0}^{n-2} f_l |\omega|^{l s_0} \big) |\omega|^{s_0-2} \nonumber\\ \label{eq:Zt_RGflow_sub-ohmic_spe}
    & - s_0(s_0-1) \int_0^{\Omega_0} \frac{\d \omega}{2 \pi} (f(\omega)-\sum_{l=0}^{n-1} f_l |\omega|^{l s_0}) |\omega|^{s_0-2} \Bigg],
\end{align}
with
\begin{align}
I^<_{s=1/n,n}+I^>_{s=1/n,n}=&\frac{-2 + n + (n+1)^2 - n \left[H(1/(n+1)) + H(1-1/(n+1)) + \pi/\sin(\pi/(n+1))\right] }{4\pi (n+1)^2},
\end{align}
where $H(x)$ is the generalized harmonic number.

\section{Numerical implementation}
In order to numerically integrate the RG equations and observe scale-invariance, we write them in a dimensionless form. From the definitions of the dimensionless couplings $\yt$ and $\yi$ in Sec.~\ref{sec:dimless}, one first infers
\begin{align}\label{eq:partial_Zt_at}
    \frac{\partial_t Z_{\tau,k}}{Z_{\tau,k}} =&2\frac{\partial_t \yt}{ \yt } - 2\frac{\partial_t v_k}{v_k},\\\label{eq:partial_alphai_ai}
    \frac{\partial_t \alpha_{i,k}}{\alpha_{i,k}} =& 2\frac{\partial_t \yi}{ \yi} + 2 - s_i  - s_i \frac{\partial_t v_k}{v_k}.
\end{align}
The integrals over frequency appearing in the RG equations are then rewritten as
\begin{align}\label{eq:int_f_omega}
    \frac{Z_x}{4 v_k} \int_{\tilde \omega}\tilde f(\tilde \omega) |\tilde \omega|^\eta =&  \sum_i \frac{ \partial_t \yi}{ \yi }\left[ 2 \yi^2 l_{s_i+\eta} -\frac{4}{s_i} \yt \yi^{2/s_i} l_{2+\eta} \right] - \frac{\partial_t v_k}{v_k} \left[2\yt^2 l_{2+\eta} + \sum_i s_i \yi^2 l_{s_i +\eta} \right]\nonumber\\
    &+ \sum_i (2 - s_i) \yi^2 l_{s_i + \eta} - 2 \bar l_\eta,
\end{align}
where $\eta$ is any exponent, $\tilde f(\tilde \omega)=kv_kf(\omega)$ is dimensionless, and the threshold functions $l_\eta$ and $\bar l_\eta$ are
\begin{align}
    l_\eta & = \int_{\tilde q, \tilde \omega} \frac{(r(y)+yr'(y))}{(y+yr(y))^2}|\tilde \omega|^\eta,\\
    \bar l_\eta & = \int_{\tilde q, \tilde \omega} \frac{r'(y)}{(1+r(y))^2}|\tilde \omega|^\eta,
\end{align}
where $y=\tilde q^2 + (\yt \tilde \omega)^2 + \sum_{i=0}^n \yi^2| \tilde \omega|^{s_i}$. We also introduce the dimensionless counterparts of the coefficients in Eq.~\eqref{eq:f_low}, i.e. $\tilde g_i=(kv_k)^{1+is}g_i$ and $\tilde h_i=(kv_k)^{1+is}h_i$. Implementing all these changes, one obtains dimensionless flow equations which can be written in matrix form as
\begin{equation}\label{eq:matrix_flow}
    A\begin{pmatrix}
    \frac{\partial_t \yo}{\yo}\\
    \vdots\\
    \frac{\partial_t \yn}{\yn}\\
    \frac{\partial_t v_k}{v_k}
    \end{pmatrix}=B
\end{equation}
where $A$ is a $(n+1)\times (n+1)$ square matrix and $B$ a vector with $n+1$ components. In the super-ohmic case ($s>1$),
\begin{align}\label{eq:dimless_superohmA}
    A=
    \begin{pmatrix}
    2 + 4\frac{v_k}{Z_x}(\frac{4}{s} \yo^{2/s} \yt l_2 - 2 \yo^2 l_s )     & -s + 4\frac{v_k}{Z_x} (2\yt^2 l_2 + s \yo^2 l_s) \\
    8(s-1) \frac{v_k}{Z_x} \yo^{2/s} ( 2 \yo^{2+2/s} l_s- s \frac{\yo^{4}}{\yt} l_{2s-2})
    & -2\yt^2 - 2 s(s-1) \frac{v_k}{Z_x} \yo^2 (2\yt^2 l_s + s \yo^2 l_{2s-2} )
\end{pmatrix},
\end{align}
and 
\begin{align}\label{eq:dimless_superohmB}
    B=\begin{pmatrix}
    s-2 + 4\frac{v_k}{Z_x} ((2-s) \yo^2 l_s - 2 \bar l_0)\\
    -2s(s-1) \frac{v_k}{Z_x} \yo^2 ((2-s) \yo^2 l_{2s-2} - 2 \bar l_{s-2})
\end{pmatrix}.
\end{align}
Similarly, for the ohmic case ($s=1$),
\begin{align}\label{eq:dimless_ohmic_A}
  A=
    \begin{pmatrix}
    -2\frac{\yo^2}{\yt} & -1 \\
    2+ 8\frac{v_k}{Z_x} (2\yo^2 \yt l_2 - \yo^2 l_1) &  -1 + 4\frac{v_k}{Z_x} (2\yt^2 l_2 + \yo^2 l_1)
\end{pmatrix},
\end{align}
and
\begin{align}\label{eq:dimless_ohmic_B}
    B=\begin{pmatrix}
    -\frac{1}{4\pi}\frac{\yo^2 }{\yt^2}\tilde g_0 \\
    -1 + 4\frac{v_k}{Z_x} ( \yo^2 l_1 -2\bar l_0 )
    \end{pmatrix}.
\end{align}
Lastly, the matrices involved in the sub-ohmic case with $s\in]1/(n+1),1/n[$ are defined by
\begin{align}\label{eq:dimless_sub-ohmic_A}
\begin{cases}
    A_{i,j} &= -\delta_{j,0} 2\yo^{2} \tilde h_{i-1} I_{s,i-1} +\delta_{i,j}2 \yi^{2}+4 \yi^{2}\frac{v_k}{Z_x} \left( \frac{4}{s_j}\yt \yj^{2/s_j} l_2 -   2 \yj^{2} l_{s_j}\right), \\
A_{i,n+1} &= s_0\yo^{2} \tilde h_{i-1} I_{s,i-1} -s_i \yi^{2} +4\yi^{2}\frac{v_k}{Z_x}\left( 2 \yt^2 l_2 + \sum_{j=0}^n s_j \yj^{2} l_{s_j} \right), \\
A_{n+1,j} &= -\frac{4}{s_j}\yj^{2/s_j}\yt - \frac{8}{s_j}\frac{v_k}{Z_x}\yj^{2/s_j} \yt\sum_{i=0}^n  s_i(s_i-1)  \yi^{2}  l_{s_i} 
 +4\frac{v_k}{Z_x}\sum_{i=0}^n s_i (s_i - 1)  \yi^{2} L_{ij},\\
A_{n+1,n+1} &= -2\yt^2 - 4\frac{v_k}{Z_x} \yt^2\sum_{i=0}^n  s_i(s_i-1) \yi^{2}  l_{s_i}
-2\frac{v_k}{Z_x}\sum_{i,j=0}^n   s_j s_i(s_i-1)  \yi^{2} L_{ij},
\end{cases}
\end{align}
where the indices $i$, $j$ run over the interval $\llbracket 0,n\rrbracket$, and 
\begin{align}\label{eq:dimless_sub-ohmic_B}
B=&\begin{pmatrix}
    \vdots\\
   (s_i - 2)\yi^{2} + \yo^{2}  \left(\tilde g_{i-1} + (2-s)\tilde h_{i-1}\right) I_{s,i-1} + 4\frac{v_k}{Z_x}\yi^{2}\sum_{j=0}^n (2-s_j) \yj^{2} l_{s_j} - 8\frac{v_k}{Z_x}\yi^{2}\bar l_0 \\
    \vdots\\
    2\frac{v_k}{Z_x}\sum_{i=0}^n  s_i (s_i - 1) \yi^{2} K_{i}-2\frac{v_k}{Z_x}\sum_{i,j=0}^n s_i (s_i - 1)(2-s_j) \yi^{2}L_{ij}\\
    \end{pmatrix}
\end{align}
where
\begin{align}
    L_{ij}&=\yj^{2} l_{s_j + s_i-2}  - \delta_{i,0}\delta_{j,0} \frac{Z_x}{4 v_k } \sum_{l=0}^{n-1} \tilde h_l J_{s,l},\\
    K_{i}&=2\bar l_{s_i-2}+\frac{Z_x}{4v_k}\delta_{i,0}\sum_{l=0}^{n-1}\tilde g_l J_{s,l},\\
    J_{s,l} &=\int_{-\infty}^{+\infty} \frac{\d \tilde \omega }{2 \pi} |\tilde \omega|^{(l+1)s-2}.
\end{align}
The coefficients $L_{ij}$ and $K_i$ must be computed as a single integral to find a convergent result (for instance, the $J_{s,l}$ are divergent on their one but make sense within $L_{ij}$ and $K_i$). Numerically, we invert the system~\eqref{eq:matrix_flow} at each time step and integrate the flow equations using a fourth-order Runge–Kutta scheme. For $s < 1/2$, the matrix $A$ becomes singular when the coefficients $\yi$ vanish. Consequently, we start the flow with a small but non zero $\tilde y_{2,k}$ in the case $1/3<s<1/3$ presented in the main text.

\section{Perturbation around the critical LL fixed point} \label{app:LL}

This section recovers the perturbative RG equations near the critical LL fixed point. This is systematically achieved by expanding in the small couplings $\yi \ll 1$, $i \in \llbracket 0, n \rrbracket$, derivatives $\partial_t \yi, \partial_t v_k \ll 1$, and distance to the critical point $x_k=K_k-(1-s/2)$, while retaining the full dependency on the speed $v_k$, which is of order $1$.

\subsection{Super-ohmic bath}
Following the process described above, the dimensionless super-ohmic RG equations~(\ref{eq:dimless_superohmA},\ref{eq:dimless_superohmB}) are expanded to give
\begin{align}
    \partial_t \yo =& \yo(s/2-1 +  4 \pi \bar C_0 K_k),\\
    \partial_t K_k  =& 2\pi s(s-1) \bar C_{s-2} K_k^2  \yo^2,
\end{align}
where the positive constants $\bar C_0$ and $\bar C_{s-2}$ are of the form
\begin{align}
    \bar C_\eta &= -\int \frac{\d \tilde q \d \tilde \omega}{4\pi^2} \frac{r'(\tilde q^2 + \tilde \omega^2)}{(1+r(\tilde q^2 + \tilde \omega^2))^2}|\tilde \omega|^\eta.
\end{align}
Using the properties $r(+\infty)=0$ and $r(0)=+\infty$ of a generic regulator $r$, one can show that $\bar C_0$ takes the universal (i.e. regulator-independent) value $1/(4\pi)$. The constant $\bar C_{s-2}$ is however not universal. With this result, expressing the RG equations in terms of $x_k$ and $\yo$ leads to
\begin{align}
    \partial_t \yo =& x_k \yo,\\
    \partial_t x_k =& 2s(s-1)(1-s/2)^2 \pi \bar C_{s-2} \yo^2,
\end{align}
which coincides with the perturbative equations derived in \cite{Majumdar2023a} for $s=1$, and in \cite{Majumdar_2023} for all $s$.

\subsection{Ohmic bath}
Repeating the same arguments for the dimensionless ohmic RG equations~(\ref{eq:dimless_ohmic_A},\ref{eq:dimless_ohmic_B}) leads to the BKT equations
\begin{align}
    &\partial_t \yo=x_k \yo,\\
    &\partial_t x_k =\frac{\bar C}{2}\yo^2,
\end{align}
with $\bar C=-\int \frac{\d \tilde q}{2\pi} \frac{r'(\tilde q^2)}{(1+r(\tilde q^2))^2}>0$.

\subsection{Sub-ohmic bath} \label{app:sub-ohmic_pert}
Expanding the generic sub-ohmic RG equations~(\ref{eq:dimless_sub-ohmic_A},\ref{eq:dimless_sub-ohmic_B}) for the coefficients $\yi$  yields 
\begin{align}
    \partial_t\yi=(K_k+s_i/2-1)\yi.
\end{align}
In the limit of small $x_k=K_k-(1-s/2)$, the scaling dimension of $\yo$ is $x_k$ while that of any $\tilde y_{i \ge 2,k}$ is $(s_i-s)/2=O(1)$, implying that the dynamics of $\yo$ is far slower than that of the couplings $\tilde y_{i \ge 2,k}$. After a transient regime, all RG trajectories should thus collapse on the plane where $\tilde y_{i\ge 2,k}=0$, resulting in a large river (or plane) effect~\cite{Bagnuls_2001a,Bagnuls_2001b}. The remaining coupling $\yo$ and $K_k$ satisfy
\begin{equation}\label{eq:y0_sub_gen}
 \partial_t \yo=x_k \yo
\end{equation}
and
\begin{equation}
    \partial_t K_k=2\pi s(s-1) \bar D_{s} K_k^2 \yo^{2},
\end{equation}
with the positive constant
\begin{align}
    \bar D_{s}=- \int \frac{\d \tilde q \d \tilde \omega}{4 \pi^2}\left[\frac{r'(\tilde q^2+\tilde \omega^2)}{(1+r(\tilde q^2+\tilde \omega^2))^2}-\frac{r'(\tilde q^2)}{(1+r(\tilde q^2))^2}\right] |\tilde \omega|^{s-2}.
\end{align}
In terms of $x_k$ and $\yo$, one arrives at
\begin{equation}\label{eq:x_sub_gen}
    \partial_t x_k=2\pi s(s-1)(1-s/2)^2 \bar D_{s} \yo^2,
\end{equation}
which, together with Eq.~\eqref{eq:y0_sub_gen}, form the usual pair of BKT equations.

\section{Perturbation around the dissipative fixed point} \label{app:TQF}

The behaviour of the dissipative fixed point (DFP) can be understood through the exponents $z$, $\eta_i$ and $\eta_\tau$ describing the scaling of couplings about the DFP as
\begin{align}
v_k \sim k^{z-1}, \qquad
\alpha_{i,k} \sim k^{-\eta_i}, \qquad
Z_{\tau,k} \sim k^{-\eta_\tau}.
\end{align}
To derive analytical expressions for these exponents, we perform a perturbative expansion of the RG equations around $\tilde y_{0,k}=1$ (so $\partial_t y_{0,k}\to 0$) and $\tilde y_{i\geq1,k}=v_k=0$. In this limit and for arbitrary $s$, one first shows that the top line of the flow equation system~\eqref{eq:matrix_flow} reduces to
\begin{equation}
-s\frac{\partial_t v_k}{v_k} = s - 2,
\end{equation}
which immediately gives the dynamical critical exponent $z = \frac{2}{s}$. Substituting $v_k \sim k^{2/s-1}$ into the definition of $\alpha_{0,k}$ in terms of $v_k$ and $\tilde y_{0,k}$ \eqref{eq:dimless} yields $\eta_0 = 0$. To determine the scaling of the higher-order couplings $\alpha_{i,k}$ appearing for $s<1/2$, one first notices from Eqs.~(\ref{eq:f_omega},\ref{eq:f_low}) that
\begin{align}
f(\omega) \sim k^{-1}, \qquad
f_i \sim k^{-1 - 2i}.
\end{align}
The sub-ohmic RG equation for $\alpha_{i,k}$ \eqref{eq:flow_sub-ohmic_1} scales as
\begin{equation}
    k^{-\eta_i}\sim k^{-\eta_i + 2/s - 1} + k^{-1-2(i-1)}.
\end{equation}
In the limit $k\to 0$, retaining the dominant contribution gives $\eta_i = 2i-1$. Similarly, the RG equation for $Z_{\tau,k}$ \eqref{eq:flow_sub-ohmic_3} scales as
\begin{equation}
    k^{-\eta_\tau} \sim k^{1-2/s} + k^0,
\end{equation}
so $\eta_\tau = \frac{2}{s} - 1$. Together, these results summarize the universal properties of the DFP for arbitrary bath exponent $s$:
\begin{align}
    z=\frac{2}{s},\quad \eta_0=0,\quad \eta_i=2i-1,\quad \eta_\tau=\frac{2}{s}-1.
\end{align}

\end{widetext}

\end{document}